\documentclass[sigconf,nonacm]{acmart}

\AtBeginDocument{%
	\providecommand\BibTeX{{%
			\normalfont B\kern-0.5em{\scshape i\kern-0.25em b}\kern-0.8em\TeX}}}

\usepackage{color, colortbl}
\definecolor{light-gray}{gray}{0.95}
\definecolor{medium-gray}{gray}{0.85}
\usepackage{amsmath,mathtools}
\usepackage{algorithm}
\usepackage{algpseudocode}
\usepackage{enumitem}
\usepackage{afterpage}
\usepackage{lscape}
\usepackage{latexsym}
\usepackage{rotating, float, caption}
\usepackage{fancyvrb}

\def\setMachines{\mathcal{M}}
\def\setUsers{\mathcal{U}}
\def\data{\mathcal{D}}
\def\techscenario{\mathbf{S}^{\text{tech}}}
\def\procscenario{\mathbf{S}^{\text{proc}}}
\def\setPosition{\mathbf{P}}
\def\EdgeTechnics{\mathbf{A}^{\text{tech}}}
\def\EdgeProc{\mathbf{A}^{\text{proc}}}
\def\Requis{\mathbf{pre}}
\def\Rewards{\mathbf{post}}
\def\controlledNodes{\mathcal{P}}
\def\connaissances{\mathcal{K}}
\def\chemin{\mathbf{C}}
\def\attackState{\mathcal{A}}
\def\startingnode{\mathbf{S}}
\def\winningnode{\mathbf{W}}

\usepackage{multirow}
\usepackage{makecell}
\usepackage{tikz}
\usetikzlibrary{arrows.meta}
\usepackage{tikz-network}
\usepackage{xr}
\usepackage{todonotes}

\begin{document}
	
	\title{URSID: Using formalism to Refine attack Scenarios for vulnerable Infrastructure Deployment}
	
	%
	\author{Pierre-Victor Besson}
	\affiliation{%
		\institution{CentraleSupélec, Inria, univ. Rennes, CNRS, IRISA}
		\city{Rennes}
		\country{France}}
	\email{pierre-victor.besson@inria.fr}
	
	\author{Valerie Viet Triem Tong}
	\affiliation{%
		\institution{CentraleSupélec, Inria, univ. Rennes, CNRS, IRISA}
		\city{Rennes}
		\country{France}}
	\email{valerie.viettriemtong@centralesupelec.fr}

	\author{Gilles Guette}
	\affiliation{%
		\institution{univ. Rennes, CNRS, Inria, IRISA}
		\city{Rennes}
		\country{France}}
	\email{gilles.guette@univ-rennes1.fr}
	
	\author{Guillaume Piolle}
	\affiliation{%
		\institution{Thales}
		\city{Rennes}
		\country{France}}
	\email{guillaume.piole@thalesgroup.com}
	
	\author{Erwan Abgrall}
	\affiliation{%
		\institution{CentraleSupélec, Inria}
		\city{Rennes}
		\country{France}}
	\email{erwan.abgrall@centralesupelec.fr}

	%
	\renewcommand{\shortauthors}{Besson et al.}
	
	\begin{abstract}
		\textbf{In this paper we propose a novel way of deploying vulnerable architectures for defense and research purposes, which aims to generate deception platforms based on the formal description of a scenario. An attack scenario is described by an attack graph in which transitions are labeled by ATT\&CK techniques or procedures. The state of the attacker is modeled as a set of secrets he acquires and a set of nodes he controls. Descriptions of a single scenario on a technical level can then be declined into several different scenarios on a procedural level, and each of these scenarios can be deployed into its own vulnerable architecture.
			To achieve this goal we introduce the notion of architecture constraints, as some procedures may only be exploited on system presenting special properties, such as having a specific operating system version. Finally, we present our deployment process for converting one of these scenarios into a vulnerable infrastructure, and offer an online proof of concept demonstration of our tool, where readers may deploy locally deploy a complete scenario inspired by the threat actor APT-29.}

	\end{abstract}
	
	%
	

	\newcommand\GG[1]{\color{blue}{#1}\color{black}}
	
	\maketitle
	
	\section{Introduction}
	The numbers of cybersecurity incidents keeps increasing every year and cybercrime is nowadays a threat to any organization. In 2021 the FBI reported a 64\% increase in losses related to cybercrime compared to 2020, and more than 4 times what it was in 2017 \cite{fbi}.
	In particular, the rise of Advanced Persistent Threats (APT) has proven problematic for companies and governments alike, leading to massive data breaches, spying and ransomware-based extortion campaigns.
	Advanced Persistent Threats are well organized stealthy threat actors, who gain unauthorized access to an information system and for extended periods of time in order to avoid detection and better reach their goals.
	In order to deal with this increasing threat, cyber defenders have an array of tools at their disposal, one of them being honeypots.

	Honeypots are deliberately insecure information systems designed and deployed with the goal of luring attackers into an environment that is controlled and monitored by the defenders. They may provide security analysts and researchers with valuable insights over the technical skills of an attacker, by studying his behavior and the means he employs to compromise the system. They may also protect existing architectures by acting as a decoy which will distract the attacker from actually sensitive systems.
	A honeypot is thus designed to be vulnerable to one or more attacks, for instance by being accessible through a brute-forceable SSH password \cite{cowrie}, or because of weaknesses in its operating system \cite{vetterl}. Honeypots may also be grouped together in the same network to form a honeynet, which acts as a decoy network.
	It is in the defender's interest to diversify the means by which his honeypots may be compromised. Indeed, the type and amount of information leaked by an attacker about his tools and skills will depend on the challenge he is tackling. Thus, deploying honeypots with diverse appearances, complexity and weaknesses may lead to defenders collecting higher quality data about threat actor attack patterns. 
	By setting up and chaining these weaknesses precisely to match known attacker patterns, the honeynet will enhance its credibility and ability to mimic an existing information system. We refer to one such advanced honeynet as a \textit{deception platform}.
	
	In order to describe these modes of operations in a more specific way, we define an attack scenario as a sequence of actions taken by an attacker in a system in order to reach his goals. For instance, an attacker may exploit a SQL vulnerability in a web server to access a user account, then exploit a CVE \cite{nvd_cve} to elevate his privileges and earn administrator rights on this system (his final goal). We consider that a deception platform is vulnerable to an attack scenario when it is designed to be compromised by performing each step of this specific scenario.
	
	
	In order to increase the amount of different scenarios a defender may offer to potential attackers, we propose an automated deception platform generation system able to generate an architecture which is vulnerable to a given attack scenario based on its description.
	
	The first contribution of this paper is a description language able to convert a unique scenario into several architectures. This language can describe the different steps of an attack scenario, and represent an architecture and the path of an attacker inside this architecture in a formal manner.
	The second contribution is our \textit{procedural refinement} algorithm, which lets us convert a single attack scenario description into several different architectures.
	The third contribution of this paper is the deployment of a vulnerable architecture resulting from our procedural refinement process using virtual machines. The necessary code is available online at \textbf{LINKTOCODEHERE} and contains instructions on how to deploy then attack the architecture, which was inspired by the mode of operation of a threat actor known as APT-29.
	
	\section{Related works}
	Two main use cases for deception platforms exist in the literature: the honeypot (as described earlier), and capture-the-flag/cyber-training generation platforms.

\paragraph{Honeypots}
Honeypots are voluntary vulnerable architectures designed to lure attackers and may be designed in several different ways depending on the needs and resources of the defenders.
Indeed, they have to balance parameters like the credibility (\textit{i.e.} how easy it is for attackers to detect that they're in a honeypot), the cost, easiness and safety of deployment. 
This is referred to as a honeypot's \textit{interaction level} and is commonly split into 2 categories:

\begin{itemize}
	\item Low interaction honeypots \cite{dionaea}\cite{tpot}\cite{conpot}, which are easy, cheap and safe to deploy. In return they are easier to detect and provide less opportunities for the attacker to demonstrate their skills. These are usually scripts running on a server giving pre-recorded answers to attackers trying to access services on this server, such as SSH or Telnet. These types of honeypots are particularly suited for analyzing botnets behavior, as bots are more likely to be fooled by low credibility systems compared to human attackers. 
	\item High interaction honeypots \cite{vetterl}\cite{iotpot}\cite{helizadirty}, which take more effort to deploy and require closer monitoring by the defender for security reasons. Indeed, attackers are given more freedom to exploit vulnerabilities and are more likely to gain full control of the machine. In return they provide attackers with a realistic system able to more extensively test their abilities. These are usually virtual machines monitored by automated logging software and with precautions taken to make sure any attacker would not be able to reuse this virtual machine as a pivot for other criminal activities. These types of honeypots are more suited to lure human attackers.
\end{itemize}

Arguably one of the most important project for low-interaction honeypots is Cowrie \cite{cowrie}. Cowrie is a SSH and Telnet Honeypot which emulates these services and a UNIX shell using python scripts, in order to make attackers think they are accessing an actual UNIX server. It can log attacker sessions, store any malware uploaded by attackers for later analysis purposes and is overall easy and safe to install. These qualities make Cowrie one of the most commonly used honeypots in literature, especially in works related to botnets analysis (such as \cite{barron2017picky}, \cite{mooselinux}). It was also updated in the past years to be capable of high-interaction emulation, albeit most of the research we found using Cowrie focus on its low interaction abilities.

For high interaction honeypots, Vetterl et al proposed Honware \cite{vetterl}, a honeypot generation platform able to quickly deploy IoT decoys based on an extensive firmware catalogue. According to \cite{vetterl}, the added flexibility and credibility of honeypots deployed with Honware compared to low-interaction alternatives makes them harder to fingerprint for attackers and better at emulating devices' vulnerabilities. Vetterl et al used Honware to deploy an ipTIME N604R wireless router and detected a previously unknown attack on this particular type of device thanks to their honeypot. 

As detailed in figure \ref{sota_table}, most of the current published projects available in the literature present limitations, such as a lack of available code, or a focus on IoT devices. They also overall lack any honeynet abilities, as these projects are focused on deploying a single honeypot at a time. Furthermore, existing projects offer limited variety in term of entry scenarios that they are vulnerable to, being able to at most change the number of machines being deployed or the specific vulnerability that has to be exploited to gain entry. We consider this variety to be an important quality for a modern \textit{deception platform} to have. We thus looked into other areas of the literature for projects with similar variety goals, which lead us to the related field of cyber-training generation platforms.

\begin{figure*}
	\begin{center}
		\begin{tabular}{ |c|c|c|c|c|c| } 
			\hline
			\textbf{Honeypot} & \textbf{Date} & \textbf{High Interaction} & \makecell{\textbf{Honeynet}} & \makecell{\textbf{Modulable entry }\\ \textbf{scenarios}} & \makecell{\textbf{Available} \\ \textbf{code}}\\ 
			\hline
			Cowrie\cite{cowrie} & 2018 & Yes & No & Limited & Yes \\
			\hline
			Dionaea\cite{dionaea} & 2013 & No & No & Limited & Yes \\
			\hline 
			Tpot\cite{tpot} & 2020 & No & No & No & Yes \\
			\hline 
			Honware\cite{vetterl} & 2019 & Yes & No & Limited & No\\
			\hline 
			Heliza\cite{helizadirty} & 2011 & Yes & No & No & No \\
			\hline
			Conpot\cite{conpot} & 2018 & No & No & No & Yes \\
			\hline 
		\end{tabular}
		\label{sota_table}
		
		\caption{\label{sota_table} Comparison of several state of the art honeypots.}
		
	\end{center}
\end{figure*}

\paragraph{CTF and cyber-training generation platforms}

Instead of generating vulnerable architectures in order to have them attacked by unknown attackers with the hopes of fooling them, researchers may instead design vulnerable architectures to be attacked by red teams, which are friendly attackers hired to test the vulnerabilities of the network. Such applications may be referred to as capture-the-flag platforms or cyber-training generation platforms, depending on their scope and use case. 

SecGen \cite{secgen} is a publicly available vulnerable virtual machine deployment tool made for educational purposes. Its goal is to be a Capture-The-Flag and student lab deployment platform, able to generate security challenges in a way that is both reusable and modulable. 
Using a combination of Vagrant \cite{vagrant}, Virtualbox \cite{virtualbox} and Puppet \cite{puppet}, it automatically generates architectures based on a xml description and populates them with vulnerabilities. SecGen also handles network connections and can randomize some of the parameters within a scenario (such as flags or passwords) in order to increase its replayability.
While the generation aspect is similar to our purposes, SecGen lacks any formal scenario representation, instead relying on the user providing a list of vulnerabilities for each machine. This makes sense in the context of generating Capture-The-Flag events but is at odd with our goal of generating different architectures based on a single scenario representation.

VulnerVAN \cite{VulnerVANTechR} is a tool aiming to generate vulnerable architectures ready to be deployed for red team training, based on a user provided description. The user describes the network architecture and an attack sequence, which is represented as a sequence of MITRE ATT\&CK techniques \cite{matrixattack}. Based on this network description and the vulnerabilities it entails, VulnerVAN calculates every possible attack paths inside this network that correspond to the attack sequence it has been given. It then generates virtual machines corresponding to the sub-network the attack path takes place in, before finally outputting a tutorial on how to exploit the architecture for the red team to follow. 

VulnerVAN fills a goal very close to our own, but has a few key differences. First of all, it does not appear to have any public release of their software yet, making its use by outside researchers for vulnerable architecture deployment purposes complicated. It also appears to be designed for red team training, which may come with different design questions than deception platforms - for instance requiring less work on the authenticity aspect, as red-team members are aware they are working on a virtual architecture. They also do not appear to model the attacker state in their formalization of attack sequences, which we do ourselves in order to represent his evolution throughout a scenario. Finally, one of the goals of our own representation was to be able to deploy several vulnerable architectures (with different vulnerabilities and configurations) for a single high level technical description of a scenario, a functionality not provided by \cite{VulnerVANTechR}.

In order to automatically generate infrastructures vulnerable to a given attack scenario, we need a formal way to express the attack scenarios and the paths that an attacker may take throughout the network. This led us to study the notion of attack graphs.

\paragraph{Attack graphs}

Attack graphs are formal structures aiming to represent one or more possible attacks on an architecture using nodes and transitions between those nodes. They differ in the literature by how they decide to represent the architecture (which depends on their use case), and most of them can be grouped in one of 3 categories:

\begin{itemize}
	\item State-based model (such as in \cite{ou}) in which a node represents the system state after the occurrence of an event and the edges represent the transitions between these states after said events happen. 
	\item Vulnerability-based model (such as in \cite{vulnegraph}), in which the graph nodes and edges illustrate the conditions, effects and interdependence of attacks and exploits in a system.
	\item Host-based model (such as in \cite{statebased}), in which nodes represent an infrastructure device and edges the accesses available between these devices. 
\end{itemize}

Among these, host-based model seemed to be the most applicable for our purposes, as information about devices can be more easily converted into virtual machine configurations than system states or a list of exploits.

Mensah et al \cite{mensah} proposes an attack graph model with a novel approach on host-based graphs. 
Indeed, the nodes in their graph hold information about not only the device the attacker is on, but also their level of privilege on said device. A node may also indicate whether it contains interesting data for the attacker, such as a password or a secret file. An attacker may move in this model by taking a transition between 2 nodes, which may have preconditions or consequences on the state of the devices and the attacker.
In order to model the fact that some transitions require that the attacker has accomplished specific actions in the network before (such as finding a password), some transitions also have \textit{triggers} attached to them. These triggers indicate that a transition can only be taken if an other specific node has been visited by the attacker beforehand.

While some aspects of this work (such as representing nodes as a list of devices and users or the concept of triggers) seemed appropriate for our problem, this formalization was ultimately designed for modeling existing (cloud) infrastructures, a goal opposite to our own goal of creating infrastructures from scratch. Their work thus contains a level of exhaustivity - in particular in how modeling a transition requires specifying all consequences and conditions of that transition - which would make it difficult to a) write scenarios manually and b) deploy different scenarios based on a single description.

Thus, while inspired by the work of Mensah et al, our contribution differs significantly in its formalism in order to fit to our generation purposes, particularly in how transitions are defined and how the attacker progresses through the network. 
We streamlined the attacker/device preconditions and consequences associated with each transition in \cite{mensah} by computing those automatically from the attacker action itself in our procedural refinement process which we describe in section \ref{scenario_proc}. This makes it easier to design a scenario from a macro scale without having to worry about every implementation detail immediately.
Furthermore, the set of actions available to the attacker used in \cite{mensah} to label their transition is not clearly defined, putting actions such as "CVE-2009-1918" or "CONNECT" on the same level.

Our goal is to deploy a vulnerable architecture based on a sequence of attack actions, and thus requires to describe precisely the set of all attack actions. We thus looked into existing nomenclatures describing cyber-attack behaviors, one of which being the MITRE ATT\&CK Matrix.

\paragraph{MITRE ATT\&CK Matrix}

The MITRE ATT\&CK Matrix \cite{matrixattack} is a cyber-adversary behavior knowledge base aiming to describe the different steps of a given attack scenario. 
It is defined on 3 levels:
\begin{itemize}
	\item \textbf{Tactics}, which represent the main goal of an action performed by an attacker. For instance, an attacker may try to gain higher-level permissions on a system or network (TA004: \textit{Privilege Escalation}).
	\item \textbf{Techniques}, which represent the means used by an attacker to reach his tactical goals. For instance, an attacker may exploit a vulnerability to elevate his privileges on a machine (T1068: \textit{Exploitation for Privilege Escalation} \footnote{We are here using the official MITRE numerotation for techniques}).
	\item \textbf{Procedures}, which describes the specific implementation details of an attack. For instance, an attacker may attempt to exploit a race condition in older linux kernel versions \cite{nvd_dirty_cow} in order to escalate his privileges on a machine.
\end{itemize}

As of the writing of this paper, the ATT\&CK matrix is comprised of 14 tactics, over 200 techniques and describes a few procedures for each technique. There is however to our knowledge no exhaustive list of procedures for a given technique, although some projects (such as \cite{bron}) provide tools to make such a connection.

Reusing this nomenclature provides us with a unique way to describe any given step of a scenario, which is needed for our application. This also makes our project more compatible with other projects using the ATT\&CK matrix \cite{bron}, or with reports relying on it to describe existing attacks or emulate new ones \cite{mitrescenario}\cite{mitrescenario3}.

	\section{Attack Scenario at the technical level}
	\label{scenario_tech}
\externaldocument{scenario_proc}
\subsection{Representation of an attack scenario}
In this work, we propose to describe an attack scenario through a directed graph of attack positions. We model the attacker through the knowledge he acquires by holding attack positions.  An attacker may progress from one attack position he holds to another if he is able to execute the attack procedure corresponding to the transition between these two attack positions. Intuitively, an attacker can progress between two different attack positions by either performing a lateral movement (the attacker compromises another account with similar privileges on an other or on the same machine) or a vertical movement (the attacker compromises another account with higher privileges on the same machine). An attacker can also remain in the same attack position but increase his knowledge by discovering information hosted on the compromised account/machine. Its progression is made possible by the use of an attack technique, itself implemented by the execution of an attack procedure.  Depending on the required level of abstraction, a scenario may be described on a technical or on a procedural level. This section describes attack scenarios at the technical level. 
The proposed formalism is generic. It can be used to describe one or many attack paths allowing to reach a targeted attack position. This formalism does not require the description of all the elements of the architecture but only the useful part of the scenario.


An \textbf{attack position} corresponds to a session of an attacker, under the identity of a user on a machine of the compromised network.  For a given architecture we define $\setMachines$ as the set of all machines available in the architecture, $\setUsers$ as the set of all users with an account on these machines and $\data$ as the data existing in the architecture. The data can be credentials, passwords, cryptographic keys, addresses, service  versions or any other data useful to the attacker to progress through the scenario..  We define an attack position by a couple $(m,u)$ where $m \in \setMachines$ is a machine of the compromised network and $u \in \setUsers$ a declared user.

Among these attack positions, a \textbf{starting position} is a position controlled by default by the attacker at the beginning of the scenario and from which he will be able to play the scenario. The starting position is usually $(AttackerInfrastructure, SuperUser)$ -we consider that the attacker has full control over his own machine(s)-, but could also be a machine from the network, in the case of an insider threat. A \textbf{winning position} is a position that is of particular interest to the attacker and which corresponds to one of his ultimate goals in the infrastructure. For instance this may correspond to an account hosting sensitive data or the admin account of a Windows Active Directory domain controller. In this formalization, we consider that an attacker can hold multiple attack positions at the same time, by compromising multiple accounts on several machines concurrently.

\paragraph{Attack techniques and procedure}
The progression of an attacker in a compromised network is made possible by the use of an attack technique, itself executed through an attack procedure according to the terminology proposed by the MITRE. In our model, this is formalized by edges between the nodes (attack positions) of a scenario graph. 

Finally, an attack scenario  at the technical level and over an architecture with  $\setMachines$  machines and $\setUsers$ declared users is a graph $\techscenario = (\setPosition, \EdgeTechnics)$ where 
\begin{itemize}
\item the set of nodes $\setPosition \subseteq \setMachines \times \setUsers$ denotes a set of attack positions. 
\item the set of edges $ \EdgeTechnics$ is  $(p_1,p_2,(\tau, \Requis, \Rewards))$ where:
\begin{itemize}
\item $p_1, p_2 \in \setPosition$ are attack positions,
\item $\tau$ is  an attack technique (or an attack procedure depending on the level of abstraction),
\item $\Requis \in \data$ are the data required to execute $t$
\item  $\Rewards \in \data$ are the data granting the attacker when he successfully completes $t$
\end{itemize}
\end{itemize}

%
%
%
We detail in section \ref{scenario_proc} how an attack scenario described at the technical level can be instantiated by attack scenarios described at the procedural level.

\begin{figure}

		\fbox{\includegraphics[scale=0.3]{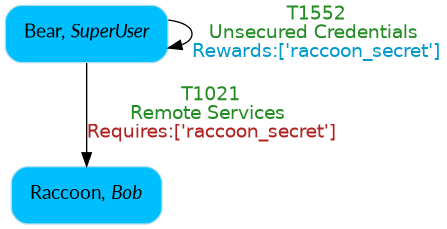}}
		\label{short}

	\caption{\label{tech_fig} Formalism associated to 2 example transitions.}
\end{figure}

To illustrate this (see figure \ref{tech_fig}), let's say an attacker acquired SuperUser Privileges on a machine named $Bear$ and is looking to move laterally to an account $Bob$ on a machine named $Raccoon$. This account $Bob$ can be accessed from outside with T1021: \textit{Remote Services} granted the attacker has acquired the secret $raccoon\_secret$ to do so. For instance this may correspond to an attacker needing to acquire a SSH key before logging through SSH on $Raccoon$. Fortunately for the attacker, Machine $Bear$ is vulnerable to T1552: \textit{Unsecured Credentials}, which rewards him with $raccoon\_secret$.
We may thus model these transitions as:
\begin{itemize}
	\item $((Bear, SuperUser), (Bear, SuperUser), (T1552,$ $\Requis=\{\}, \Rewards={raccoon\_secret}))$ for the unsecured credential transition.
	\item $((Bear, SuperUser), (Raccoon, Bob), (T1021,$ 
	$\Requis$=\\$\{raccoon\_secret\}$ $\Rewards=\{\}))$ for the lateral movement transition. 
\end{itemize} 


\paragraph{Attacker progression}
For a given scenario $\techscenario = (\setPosition, \EdgeTechnics)$, we model the attacker's progression as {\bf an attack state} defined by $\attackState = (\controlledNodes, \connaissances$), where $\controlledNodes \subseteq \setPosition$ is the set of all attack positions he controls, and $\connaissances \subseteq \data$ the data he acquired. In our model, we consider that an attacker may not lose the control of an attack position after he acquires it once. Indeed, we suppose that once he manages to open a session on a given machine by taking a sequence of transitions, the attacker is able to get back to that attack position by applying the same sequence, or by using a persistence tactic \cite{matrixattack}. While this hypothesis may exclude some destructive attacks or unstable exploits, these represent a small proportion of attacks in practice and considering node acquisition to be permanent will make it easier to traverse the graph in later algorithmic treatments.

The attacker progresses in the scenario by controlling new attack positions or increasing his knowledge on the architecture.  If an attacker controls an attack position $p_1$  
and if in the attack scenario there exists a position $p_2$, a technique $\tau$, two data sets $\Requis$ and $\Rewards$ such that the attacker already knows $\Requis$  
then the attacker can progress from $p_1$ to $p_2$ by applying $t$. If the execution of $\tau$ succeeds, the attacker is rewarded with $\Rewards$.

An attack path is thus a finite sequence of attack states $ \chemin = \attackState_{0} \stackrel{\tau_1}{\rightarrow} \attackState_{1} \stackrel{\tau_2}{\rightarrow}  \ldots  \stackrel{\tau_n}{\rightarrow}\attackState_{n}$. An attacker can progress from a state $\attackState_{i} = (\controlledNodes_{i}, \connaissances_{i})$ to a state $\attackState_{i+1} = (\controlledNodes_{i+1}, \connaissances_{i+1})$ by applying a technique $\tau$ in a scenario $\techscenario = (\setPosition, \EdgeTechnics)$ if
 
\begin{itemize}
\item $(p_1,p_2,(\tau, \Requis, \Rewards)) \in \EdgeTechnics$
\item $ p_i \in \controlledNodes_{i}$
\item $\Requis \in \connaissances_{i}$
\item $\controlledNodes_{i+1} = \controlledNodes_{i} \cup \{p_2\}$ 
\item $\connaissances{i+1} = \connaissances_{i} \cup \Rewards$ 
\end{itemize}

An attack path $ \chemin = \attackState_{0} \stackrel{\tau_1}{\rightarrow} \attackState_{1} \stackrel{\tau_2}{\rightarrow}  \ldots  \stackrel{\tau_n}{\rightarrow}\attackState_{n}$ is said executable if the attacker can progress from $ \attackState_{0} $ to $ \attackState_{n} $. 

%
%
%
%
%
%
%
%
%

\paragraph {Winnable scenario} 
Among the scenarios which can be described with this formalism, we distinguish the so-called {\bf winnable  scenarios} in which there is, at least, an executable attack path $ \chemin = \attackState_{0} \stackrel{t_1}{\rightarrow} \attackState_{1} \stackrel{t_2}{\rightarrow}  \ldots  \attackState_{n}$, a starting attack position $p_0$ such that $p_0 \in \controlledNodes_{0}$  and a winning position $p_w$ such that $p_w \in \controlledNodes_{n}$.

\subsection{Relevant attack tactics} 
As previously discussed, the techniques used here correspond to a specific tactic.  Among the $14$ tactics presented by the MITRE Att\&CK Matrix in \cite{matrixattack}, some are out of the scope of our study which focuses on the specification of an attack scenario on a vulnerable architecture. 
We only consider tactics that may lead to changes in an attack state, by giving the attacker new attack positions or secrets. For instance applying TA0004: \textit{Privilege Escalation} usually leads to a new attack position for the attacker.  We do not take into account here the actions of the attacker made outside this architecture. For instance, the tactic TA0043: \textit{Reconnaissance} comprises of techniques used by attackers to gather information about an architecture and its users before attempting to infiltrate it. We thus considered this tactic and its relevant technics and procedures to be outside the scope of our project.

%
%
%

Table ~\ref{tactic_table} lists all tactics and shows which one we consider relevant for our purposes. The lines on grey background are implemented in our example scenario.

\begin{center}
		\begin{table}
	\begin{tabular}{ |c|c|c| } 
		\hline
		MITRE Tactic Name & Relevant & \makecell{Present in \\ example scenario}\\ 
		\hline
		Reconnaissance & No & No \\ 
		Resource Development & No & No \\ 
		\rowcolor{medium-gray}
		Initial Access & Yes & Yes \\
		Execution & No & No \\
		Persistence & No & No \\
		\rowcolor{medium-gray}
		Privilege Escalation & Yes & Yes \\
		Defense Evasion & No & No \\
		\rowcolor{medium-gray}
		Credential Access & Yes & Yes \\
		Discovery & No & No \\
		\rowcolor{medium-gray}
		Lateral Movement & Yes & Yes \\
		\rowcolor{medium-gray}
		Collection & Yes & Yes \\
		Command and Control & No & No \\
		Exfiltration & No & No \\
		Impact & No & No \\
		\hline
	\end{tabular}
	\label{tactic_table}
	\caption{\label{tactic_table} Relevant attack tactics for our approach.}
	\end{table}
\end{center}

\subsection{Case study} 
To illustrate this article, we rely on the MITRE Adversary Emulation Plans \cite{mitrescenario}, which outlines the behavior of persistent threat groups mapped to ATT$\&$CK. Our scenario is directly inspired from the “smash and grab" attack“ of Day 1 of the MITRE APT29 Adversary Emulation Plan. In this scenario, the attacker first gets access to a machine and escalates his privileges on it. He then acquires passwords and files from this position before performing lateral movement to an other machine, which he will also loot and exfiltrate files from. The Emulation Plan details every technique used by the attackers during this scenario. Our own scenario thus follows the same outline, with a few differences:
\begin{itemize}
	\item We only chose techniques which belong to ATT\&CK tactics we consider relevant, as described in table \ref{tactic_table}.
	\item Some of the techniques described belong to tactics (like Extraction or Discovery) we have chosen not to represent, and were therefore ignored, as shown in table \ref{tactic_table}.
	\item The original MITRE scenario only describes 2 machines. We added 2 more in order to give the attackers more options after lateral movement and to diversify his possible attack paths. This could thus let us see which of the machines are more attractive to an attacker, and whether this correlates with their privilege escalation technique or which type of files they contain (in a similar way to \cite{barron2017picky}).
\end{itemize}

Our scenario is played out on 4 machines, named \textit{Bear, Raccoon, Badger and Skunk}. Each machine has 2 users, with one of them being a $SuperUser$ with elevated privileges. There is also an additional node $(AttackerInfrastructure, SuperUser)$ representing the starting position of the attacker. The goal of the attacker is to reach the winning position $(Skunk, root)$. 

At a technical level, our scenario can be described by the graph  $\techscenario_{e} = (\setPosition_{e}, \EdgeTechnics_{e})$ where $\setPosition_{e} = {(p_i)_{i \in [1,9]}} $,  $\EdgeTechnics_{e}=  {^(\tau_i)_{i \in [1,16]}}$, $\startingnode_{e} = \{p_1\}$ and $\winningnode_{e} = \{p_9\}$. An automatically generated graphical representation of this scenario detailing the values of ${(p_i)_{i \in [1,9]}}$ and ${^(\tau_i)_{i \in [1,16]}}$ is available in figure \ref{scenario_fig}.

\begin{figure*}
\begin{center}
	\fbox{\includegraphics[scale=0.3]{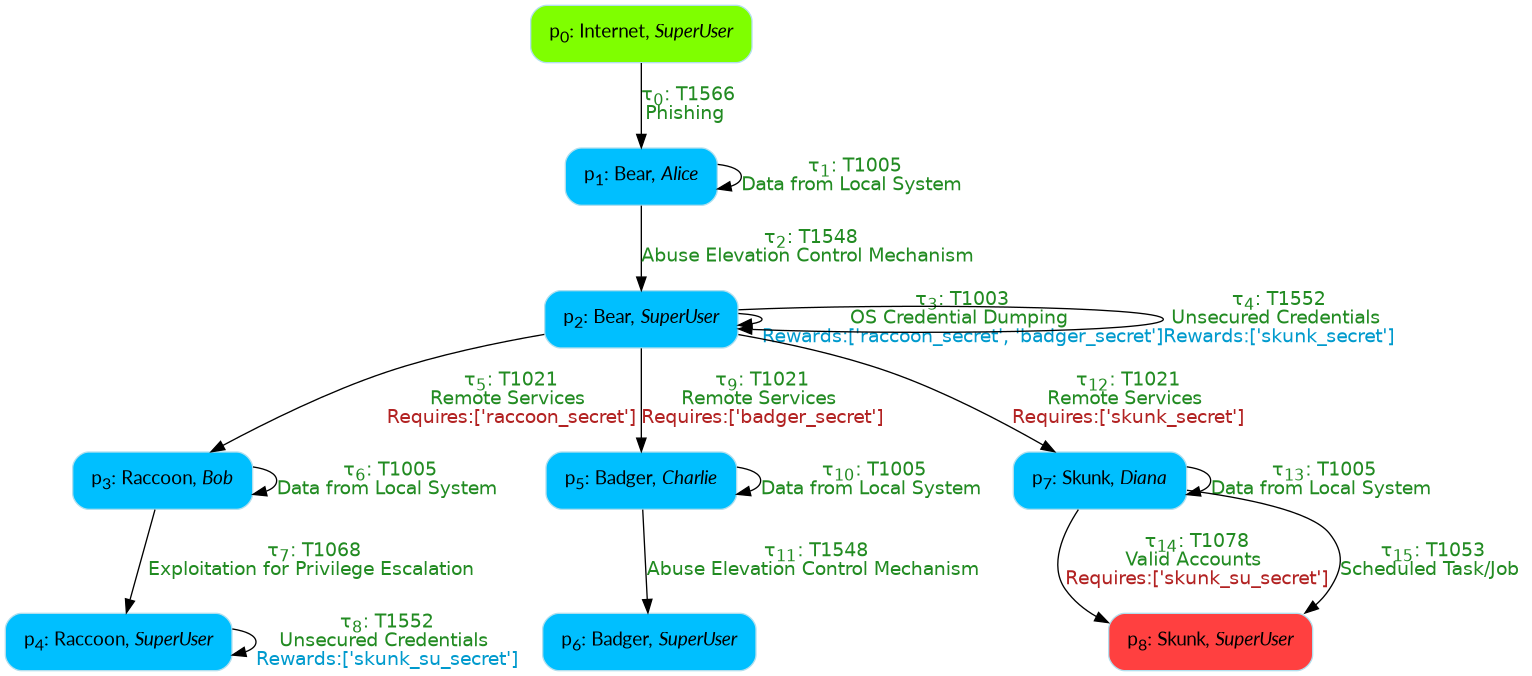}}
	\label{scenario_fig}
\end{center}
\caption{\label{scenario_fig} Automatically generated graphical representation of our example scenario.}
\end{figure*}

The attacker has 2 main paths to the winning position  here. He must always start by gaining an initial access to machine $Bear$, perform a Privilege Escalation technique to acquire SuperUser privileges, then harvest secrets available on the machine. Then he may go directly to machine $Skunk$ and perform T1053: \textit{Scheduled Task/Job} to get to the winning position $(Skunk, SuperUser)$, or he may go to machine $Raccoon$ first to acquire the credentials necessary to perform T1078: \textit{Valid Accounts} to achieve the same goal: the winning position $(Skunk, SuperUser)$.
It has to be noted that machine $Badger$ is unnecessary for both of these winning attack paths here, acting as a dead end to diversify the paths an attacker may take in our network. 

%
%
%

	\section{From attack techniques to attack procedures}
	
\label{scenario_proc}

In the previous section, we proposed an attack scenario representation through a directed graph of the different attack positions that an attacker can hold on the target system. This first modeling is done at the technical level and allows for the description of a relevant subset of attacker actions in the network. This high-level of description, while convenient to describe scenarios based on the attacker's choice of ATT\&CK techniques, is by design not sufficient to grasp the implementation details necessary to actually deploy the vulnerable system. In this work, we are interested in the automatic generation of attack scenarios and vulnerable architectures allowing to play these scenarios. The vulnerable architecture is defined by the machines (including the version of the operating system, the version of the installed services, the files and their contents) that compose it and the declared users (names, passwords and privileges) on these machines. To reach this level of detail we propose to refine the description of an attack scenario at the level of attack procedures.

For each technique $t$ allowing to move from an attack position $(u_1,m_1)$ to a position $(u_2,m_2)$, the choice of an attack procedure $p$ instantiating $t$ induces architectural constraints on machines $m_1$ and $m_2$. We consider here that an attack procedure \textbf{instantiates} an attack technique when it is a possible practical implementation of this technique. For instance, MITRE ATT\&CK \cite{matrixattackLSASS} details a procedure that can be used with PowerShell to inject into the process memory of the Local Security Authority Subsystem Service {\tt lsass.exe} in order to harvest credentials. In this example, the technique used is {\it OS Credential Dumping} that aims to dump credentials to obtain account login and credential material. This same technique could also have been implemented  by extracting credential material from the Security Account Manager (SAM) database \cite{matrixattackSAM}. Targeting the SAM database or the {\tt lsass} process are two attack procedure instantiating the same attack technique.  These two procedures however only makes sense on a windows machine. The choice of one procedure to instantiate a technique implies constraints on the machines of the architecture on which the scenario will be played. 

\paragraph{Constraint}
A \textbf{constraint} $\mathcal C$ is related to a specific machine $m$ and is denoted by $\mathcal C (m)$. A single constraint $\mathcal C$ is comprised of several subconstraints, including Operating System constraints (OS), Account constraints, Software constraints and File constraints.
An OS constraint corresponds to the choice of an OS type (such as Windows, Windows server, Ubuntu, Debian, Fedora, etc. ) and a version or a set of versions for this OS. We consider that a single machine may run only one OS at a time, as virtual machines and containers may be modeled as separate machines. 
An account constraint details the name, group, privilege and credentials of users which have to be added on a machine for the procedure to be available. 
A software constraint specifies a list of software (name, version and port if applicable) that have to be installed on a machine. These software are installed for every user by default.   
Finally, a file constraint details information about the files (name, path, permissions, content) that have to be added or modified on a machine. This information includes the path of the file on the resulting machine, the permissions associated to this file and its content. 
Formally, we define $\mathcal C (m)$ as:

$$
\mathcal C (m) = \left\{
\begin{array}{ll}
	OS & type, version \\
	Account & \text{list of } (name, group, privilege, credentials)\\
	Software & \text{list of }(software, version, port) \\
	Files & \text{list of }(path, permissions, content) \\
	
\end{array}
\right.
$$

\paragraph{Compatible procedures}
The structure of a technical scenario (in term of its secret requirements and rewards) may induce additional constraints on which procedures are eligible to refine a given technique. For instance, some procedures (such as connecting to a remote account by knowing its SSH password) will always require the attacker to have gathered previous information to be applied. As such, these procedures should only be picked when the associated transition in the technical graph has at least one secret requirement, guaranteeing that this secret will have been generated elsewhere in the scenario. Blindly picking these procedures for transitions that do not reward or require secrets on a technical level may lead to scenarios that are in practice impossible to complete.

As such, in addition to their architecture constraints a procedure $p$ may have \textbf{secret preconditions} $p_{pre}$. These preconditions indicate which kind of secret (ssh keys, passwords...) this procedure may reward or require, alongside how many of these secrets it may handle. When refining the scenario, procedures will only be picked if they match these preconditions. Then, if one of the secrets related to that transition has already been generated ({\it i.e.} its type and value have been set), secret preconditions guarantee that any new procedure trying to manage the same secret will be equipped to handle that specific type of secret.

$$
p_{pre} = \left\{
\begin{array}{ll}
	Requires & type, number \\
	Rewards & type, number\\
\end{array}
\right.
$$

For a given transition $\tau$ and the procedures $\{ P_{\tau} \} $ instantiating its technique, we thus define $Comp(\tau)$ as the subset of $\{P_{\tau}\}$ which fits the preconditions described above. $Comp(\tau)$ will represent the set of all procedures that are eligible to refine this transition in this particular scenario. 
For instance, let's consider $\tau = ((Bear, SuperUser), (Raccoon, Bob), (T1021, \Requis$=$\{raccoon\_secret\}$ $\Rewards=\{\}))$ a transition associated to technique T1021 = Remote Services, in which the attacker logs into a machine specifically designed to accept remote connections \cite{matrixattackRemoteServices}. This could be done by either reusing a previously found password to access an account through SSH ($P_1)$, or by brute-forcing the SSH password of that account ($P_2$), giving us $\{ P_{\tau} \} = \{P_1, P_2\}$. Here, $P_2$ does not require or reward any secrets, while $P_1$ requires the attacker to know of a previous password secret. Since $\Requis$ $\neq$ $\emptyset$ here, only $P_1$ is eligible for this specific transition, and therefore $Comp(t) = \{P_1\}$

\paragraph{Use of the CPE format} 
We rely when available on the CPE naming format \cite{nvd_cpe} for some of our constraints. This format is a nomenclature defined as "a structure naming scheme for information technology systems, software, and packages", and aims to provide a unique way of describing any kind of hardware or software configuration. 
In particular, this database provides unique denominations for a wide range of operating systems and software, related to their vendor name, product name and versions. Thus, when available we name the $type$ entry of our OS constraints and the $software$ entry of our Software constraints, by using a combination of their associated vendor name and product.
We chose to be compatible with this format because it is also used in the National Vulnerability Database project \cite{nvd_cve} that provides example of configurations vulnerable to specific CVE in the CPE format. Since "exploiting a specific CVE" is a procedure, this lets us gather OS and software constraints for these procedures more easily, and may even let us reuse some projects that have attempted to automatize this process \cite{bron}.

\paragraph{Refinement from techniques to procedure} 
A procedural level scenario follows the same formalism as a technical level scenario, except that the edge sets are labeled by attack procedures rather than by techniques. The constraints generated by the choice of procedures make it possible to specify the architecture when these constraints are not inconsistent between them. We propose a \textbf{backtracking refinement algorithm} \ref{alg:backtracking} to generate an architecture that can play a scenario described at the procedure level. This algorithm walks through the scenario graph and try to replace each attack technique labeling an edge in the scenario by an attack procedure instantiating the technique as long as the constraints on the architecture stay coherent. 
The rules for adding constraints are given in figure \ref{fig:table_and} . Intuitively, two constraints on the same machine add up if they have different software, users, credentials or if one of the two constraints is more restrictive than the other in terms of OS and software versions. Otherwise, they are considered to be inconsistent.

To this aim, we need to determine whether individual subconstraints on comparable parameters are compatible, and if so, the result of their combination. We therefore introduce the $\bowtie$ operator, applicable to the various types of subconstraints already defined. Given two subconstraints $Sc_1$ and $Sc_2$, $Sc_1$ $\bowtie$ $Sc_2$ returns a new subconstraint $Sc_3$, which can be either $\bot$, which is the unsatisfiable subconstraint and a conclusion of incompatibility, or a valid combination of $Sc_1$ and $Sc_2$. Of course, the details of this combination depend on the nature of the operands. Formally, there is a different $\bowtie$ operator for each type of subconstraint. For the sake of clarity, we choose to use the same symbol, while always making sure that both operands are of the same type.
Let us take OS constraints as an example. An OS constraint is a tuple of a type subconstraint and a version subconstraint. Quite intuitively, the $\bowtie$ operator applied to "type=Linux" and "type=Debian" would return "type=Debian", while on "type=Linux" and "type=Windows" it would return $\bot$. This version of the operator cannot be defined in a very formal way. When it comes to version numbers, however, it behaves much like a linear constraint solver, under the hypothesis that version numbers can always be compared in a safe manner. In a comparable fashion, we define the semantics of the $\bowtie$ operator on other types of subconstraints.
Based on this combination operator, we are then able to define a fusion operator $\sqcup$ over constraint sets, and to determine whether the result remains satisfiable or not. Given two constraints $C_1$ and $C_2$, $C_1 \sqcup C_2$ returns a new constraint $C_3$, which can be either the unsatisfiable constraint $\bot$ or a valid combination of $C_1$ and $C_2$.
Finally, the set of all constraints over the resulting architecture is represented as a dictionary of constraints $C$ containing entries for every machine $m$. This dictionary will be filled and edited throughout the course of our backtracking algorithm, and is considered to be unsatisfiable (C = $\bot$) if any of its entries are unsatisfiable..

\begin{figure*}
	\begin{center}
		\begin{tabular}{ |c|c|c| } 
			\hline
			\textbf{Subconstraint type} &  \textbf{$Sc^{type}_{i}$} & \textbf{$Sc^{type}_{1}\sqcup Sc^{type}_{2}$} \\ 
			\hline 
			OS & $type_i, version_i$ & \makecell*[{{p{11cm}}}]{$((type_1 \bowtie type_2), (version_1 \bowtie version_2))$} \\
			
			\hline 
			Account &  \makecell[c]{$name_i, group_i, priv_i, $\\$ cred_i$} & \makecell*[{{p{11cm}}}]{$(name_1, group_1, priv_1, cred_1) \sqcup (name_2, group_2, priv_2, cred_2)$ if $name_1 \neq name_2$ \\ $(name_1, (group_1 \bowtie group_2), (priv_1 \bowtie priv_2), (cred_1 \bowtie cred_2))$ if $name_1 = name_2$} \\  
			\hline 
			Software & $type_i, version_i, port_i$ & \makecell*[{{p{11cm}}}]{$(type_1, version_1, port_1) \sqcup (type_2, version_2, port_2)$ if $(type_1 \neq type_2 \land port_1 \neq port_2)$\\ $(type_1, (version_1 \bowtie version_2), (port_1 \bowtie port_2))$ if $type_1 = type_2$
			} \\
			
			\hline 
			Files & $path_i, perm_i, content_i$ & \makecell*[{{p{11cm}}}]{$(path_1, (perm_1 \bowtie perm_2), (content_1 \bowtie content_2))$ if $path_1 \neq path_2$ \\ $(path_1, perm_1, content_1) \sqcup (path_2, perm_2, content_2)$ if $ path_1 = path_2$ } \\
			\hline

		\end{tabular}
	\end{center}
	\caption{Constraint combination rules for 2 procedures}\label{fig:table_and}
\end{figure*}

\paragraph{Notes on secret generation}

Secrets throughout the scenario refinement process are generated on a first come first served basis.
Each time a procedure is picked that requires or rewards secrets, a value for this secret will be generated according to its type (such as ssh key or plaintext password) if it hasn't been generated already. The name of this secret alongside its type and value will then be stored in a global dictionary, so that any future procedure reusing this secret may fetch its value.
In particular this is relevant for procedures which need to set file contents or account credentials to values corresponding to generated secrets - for instance a procedure which leaks a password of an account in a text file.

\paragraph{Resulting Configuration file}
The virtual machine configuration file resulting from all the procedure choices and their associated constraints is a list of machines and their configuration. For each machine $m$, the corresponding entry in the configuration file $C[m]$ is initialized with generic values and constantly updated as constraints are added through algorithm \ref{alg:backtracking}. This configuration file contains enough information about the machines comprising the network - which users to add, which software and OS to install, which files to add and create - to be used for virtual machine deployment, a process we will describe in section \ref{generation}. 
We also make sure that the set of transitions $\EdgeTechnics$ is sorted to put transitions requiring or reward secrets first, in order to optimize computation times. This is done because secret types are one of the more commons causes of architecture incompatibilities. 
After the refinement process some values (such as account credentials or privilege) may have no constraints attached to them, in which case they are set to default values. For instance, passwords that have no constraints are set to a random hash value, and when not assigned a group (a group sharing the same name as its user is created).

\begin{algorithm}
	\caption[Backtracking refinement algorithm for a scenario $\techscenario = (\setPosition, \EdgeTechnics)$]{Backtracking refinement algorithm for a scenario $\techscenario = (\setPosition, \EdgeTechnics)$ \protect \footnotemark} \label{alg:backtracking}
	\begin{algorithmic}
		
		\Require $\techscenario = (\setPosition, \EdgeTechnics)$ a scenario described at the technical level ($\EdgeTechnics$ are attack techniques)
		\Require $C$ a set of architectural constraints.
		\Require $\{ Proc_{\tau_i} \}_i $ a list of procedures instantiating each $\tau_i$ 
		\State $\procscenario := (\setPosition, \emptyset)$
		\State $L_t := \EdgeTechnics$ the list of all transitions to refine, sorted to prioritize transitions requiring or rewarding secrets.
		
		\Ensure $\procscenario = (\setPosition, \EdgeProc)$  a scenario described at the procedural level ($\EdgeProc$ are attack procedures)
		and  $\procscenario \vartriangleright  \techscenario$
		
		\Function {backtrack} {$\techscenario, \procscenario, L_{t}, \{ P_{\tau_i} \}_i , C$}
		\If{ $L_{t} = \emptyset$ }
		\Return True, $C$

			\Else{
				\State Sort $L_{t}$, putting transitions requiring or rewarding secrets first.
				\State $t = L_t[0] =  ((u1,m1)$,$(u2,m2)$,$(\tau, \Requis$, $\Rewards))$
				\If{$\{ P_{\tau} \} = \emptyset$}
					\Return False, $\emptyset$	
				\EndIf
				\State Get the set of all compatible procedures $\{ Comp(t) \}_i $
				\ForAll{$\{proc \in Comp(t) \} $}
					\State $C_{new} = C$
					\State $C_{new}[m1] = C_{new}[m1] \sqcup proc[m1]$
					\State $C_{new}[m2] = C_{new}[m2] \sqcup proc[m2]$
					\State$\procscenario_{new} = (P, \EdgeProc + {proc})$
					\If{$C_{new} \nvdash \bot$}
						\State $L_{tnew} = L_t - \{t\}$
						\State isValid, finalC = backtrack($\techscenario, \procscenario$, $L_{tnew}, \{ proc_{t_i} \}_i, C_{new} $)
						\If{isValid} 
							\Return True, finalC
						\EndIf
					\EndIf 
				\EndFor 
			
						\Return False, C \Comment{No valid procedure was found}
			}
			\EndIf 
	
		\EndFunction
	\end{algorithmic}
\end{algorithm}
\footnotetext{We surcharge the "=" operator here to also indicate deep copies of elements (such as in $C_{new} = C$).}

\paragraph{Example}

We will now illustrate our procedural refinement algorithm with a short example shown in figure \ref{fig:refinement_example}, which is a subset of our example scenario described in \ref{scenario_tech}.
This scenario consists of 3 transitions, and requires the attacker to apply T1548: \textit{Abuse Elevation Control Mechanism} on machine \textit{Bear} to elevate their privileges. They then run T1552: \textit{Unsecured Credentials} to acquire the secret \textit{skunk\_secret}, before finally reusing that secret to gain remote access to machine Skunk as account \textit{Diana} using T1021: \textit{Remote Services}.

We listed 2 procedures for each technique, which we detail in annex \ref{fig:new_table_example}. These procedures have various constraints (operating system, secret preconditions...) which will come into play during the refinement process. 
Our algorithm first picks a procedure at random to refine the transition associated to T1548, and lands on the procedure "CVE to bypasse UAC", which lets attackers achieve remote code execution thanks to a software vulnerability \cite{easyinstallcve}. This induces constraints on the resulting architecture, in particular that machine \textit{Bear} must run on a Windows.
Then T1552 is refined, and the algorithm attempts to pick the "Passwords in bash history" procedure and computes the resulting constraints. However, since machine \textit{Bear} must be a Windows and this procedure has OS constraints related to Linux, the resulting architecture is invalid and the algorithm backtracks. Thus, "Passwords in text file" is picked, and its resulting constraints are computed and found compatible. This procedure also generates secret skunk\_secret, which as indicated by the secret precondition is now of type Password.
Finally, T1021 is refined. However, "SSH Access from key" requires a SSH key as a secret, which does not fit with the type of the secret we already generated. It is therefore eliminated from the refinement process, leading the algorithm to land on "RDP Access from password" as a procedure, implying constraints on machine \textit{Skunk}.

\begin{figure}
	\begin{center}
		\fbox{\includegraphics[scale=0.3]{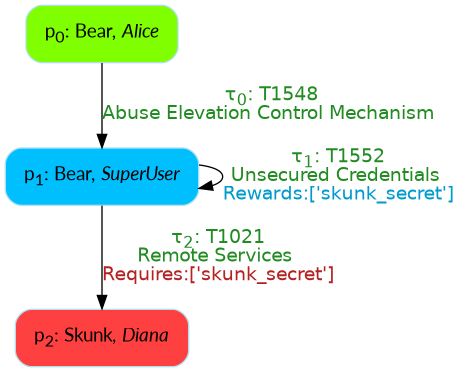}}
		\label{fig:refinement_example}
	\end{center}
	\caption{\label{fig:refinement_example} Illustratory excerpt from our example scenario.}
\end{figure}

\paragraph{Application} We implemented a few procedure constraints for each technique in our example scenario. The summary of every procedure chosen and their associated constraints is presented in table ~\ref{procedure_choice_table}.

\begin{figure*}
	\begin{center}
		\begin{tabular}{ |c|c|c| } 
			\hline
			\textbf{Technique} &  \textbf{Procedure} & \textbf{Constraint notes}\\ 
			\hline
			\multirow{2}{3em}{\textbf{T1566}} & Linux Payload \ &  Requires Linux\\
			& Windows Payload & Requires Windows \\
			\hline
			\multirow{2}{3em}{\textbf{T1005}} & Business File Corpus \ &  Adds Files \\
			& Student File Corpus & Adds Files \\
			\hline
			\multirow{2}{3em}{\textbf{T1548}} & Bad Sudo Config \ &  Requires Linux, edits sudoers\\
			& CVE to disable UAC & Requires Windows, installs flawed software \\
			\hline
			\multirow{2}{3em}{\textbf{T1003}} & Crackable Password in /etc/shadow \ &  Requires Linux, root, a crackable hash, rewards a plaintext password\\
			& SAM dump & Requires Windows, SYSTEM, rewards a hashed password \\
			\hline
			\multirow{3}{3em}{\textbf{T1552}} & Password in bash history \ &  Requires Linux, append a secret to the bash history file, rewards a plaintext password\\
			& Private keys in .ssh & Requires Linux, creates files using a secret, rewards SSH keys \\
			& Passwords in text file & Creates a file using a secret, rewards a plaintext password \\
			\hline
			\multirow{3}{3em}{\textbf{T1021}} & Weak SSH password \ &  Requires Linux, ssh service on port 22, sshd\_config allowing password authentication\\
			& SSH access from key & Requires Linux, a ssh service on port 22, authorized\_keys to be edited using a secret \\
			& SSH access from password & Requires Linux, a ssh service on port 22, ssh\_config to allow password connections \\
			& RDP from Password & Requires Windows, a RDP service on port 3389, Windows registry + user password edits. \\
			\hline
			\multirow{2}{3em}{\textbf{T1068}} & Dirty COW \ &  Requires specific Linux version\\
			& Vulnerable sudo version & Requires Linux, specific sudo package, edit sudoers \\
			\hline
			\multirow{2}{3em}{\textbf{T1053}} & Cronjob with bad permissions \ &  Requires linux, files to be created and edited\\
			& Scheduled Job with bad permissions & Requires Windows, files to be created and edited \\
			\hline
			\multirow{2}{3em}{\textbf{T1078}} & Password from secret \ &  Requires the user password to match a secret\\
			& Weak Password & Requires the user password to be easy to brute-force.\\
			\hline 
		\end{tabular}
		\label{procedure_choice_table}
	\end{center}
	\caption{\label{procedure_choice_table} Available procedures for each technique in our example scenario.}
\end{figure*}

We then ran our backtracking algorithm on our scenario, with at least 2 different procedures for each technique, 4 possible versions of Windows and 10 possible versions of Linux to deploy. Some remarks on the process are as follows: 

\begin{itemize}
	\item Having a procedure decide whether the machine runs on Windows or Linux quickly limits the options for the next procedures on the same machine.
	\item The choice of procedure for technique T1552: \textit{Unsecured Credentials} has consequences on the techniques available for T1021: \textit{Remote Services} between $(Bear, SuperUser)$ and $(Skunk, Diana)$. Indeed if T1552 rewards a SSH key to the attacker, this instance of T1021 cannot pick a procedure requiring a password (for instance a RDP connection).
	\item The procedure chosen for technique T1021: \textit{Remote Services} between $(Bear, SuperUser)$ and $(Racoon, Bob)$ can only be a procedure fit to require a secret, and thus cannot be "Weak SSH Password".
	\item Since T1078: \textit{Valid Accounts} procedures only accept passwords, T1552: \textit{Unsecured Credentials} between \textit{(Raccoon, SuperUser)} and itself may not give a SSH key as a reward.
	\item The output architectures have varying numbers of Windows and Linux Operating System on them. Machine $Raccoon$ must be on a Linux (because of our available procedures for T1068), but $Skunk$, $Badger$ and $Bear$ can be either.
	\item The number of possible resulting architectures exponentially grows as we increase the number of machines deployed, procedures implemented or amount of OS we are able to deploy. We computed it as 334016 for this scenario, although this number is very much inflated by the number of valid OS versions we chose, as well as each machine having 2 different OS-agnostic file corpuses available. This is however not an issue for complexity, as our refinement backtracking algorithm only picks one of these possible resulting architectures and is not required to go through all possibilities.
\end{itemize}

\begin{figure*}
	\begin{center}
		\begin{tabular}{ |c|c|c|c|c|c|c| } 
			\hline
			\textbf{Bear} &  \textbf{Raccoon} & \textbf{Badger} & \textbf{Skunk} & \textbf{$N$} & \textbf{$N_{OS}$} & \textbf{$N_{OS,F}$} \\ 
			\hline
			Windows & Windows & Windows & Windows & 0 & 0 & 0 \\
			\hline
			Windows & Windows & Windows & Linux & 0 & 0 & 0 \\
			\hline
			Windows & Windows & Linux & Windows & 0 & 0 & 0 \\
			\hline
			Windows & Windows & Linux & Linux & 0 & 0 & 0 \\
			\hline
			Windows & Linux & Windows & Windows & 4 & 544 & 8704 \\
			\hline
			Windows & Linux & Windows & Linux & 4 & 1360 & 21760 \\
			\hline
			Windows & Linux & Linux & Windows & 4 & 544 & 8704 \\
			\hline
			Windows & Linux & Linux & Linux & 4 & 1360 & 21760 \\
			\hline
			Linux & Windows & Windows & Windows & 0 & 0 & 0 \\
			\hline
			Linux & Windows & Windows & Linux & 0 & 0 & 0 \\
			\hline
			Linux & Windows & Linux & Windows & 0 & 0 & 0 \\
			\hline
			Linux & Windows & Linux & Linux & 0 & 0 & 0 \\
			\hline
			Linux & Linux & Windows & Windows & 8 & 1088 & 17408 \\
			\hline
			Linux & Linux & Windows & Linux & 12 & 4080 & 48960 \\
			\hline
			Linux & Linux & Linux & Windows & 8 & 2720 & 43520 \\
			\hline
			Linux & Linux & Linux & Linux & 12 & 10200 & 163200 \\
			\hline

		\end{tabular}
		\label{resulting_scenarios_table}
	\end{center}
	\caption{\label{resulting_scenarios_table} Number of generated architectures available for each OS combination for our scenario. $N$ groups architectures which only differ by their OS version or file corpuses. $N_{OS}$ groups architectures which only differ by their choice of file corpuses. $N_{OS,F}$ represents the total number of different deployable architectures. }
\end{figure*}

	\section{Vulnerable architecture generation} \label{generation}
	In the previous section we discussed the procedural refinement process, which lets us decline a single scenario described on a technical level into several possible scenarios described on a procedural level. This refined description states architecture constraints, such as the required operating systems, software, files that need to be created or edited and credential values. This description is sufficient to deploy the corresponding architecture, using Vagrant \cite{vagrant} and Ansible \cite{ansible} to instantiate and populate the virtual machines, and VirtualBox \cite{virtualbox} to deploy them.

As a proof of concept, we have refined the scenario based on our APT-29 example, and implemented every relevant procedure for that specific refinement. This scenario is comprised of 4 Linux machines and is vulnerable to some of the procedures described in table \ref{procedure_choice_table}, letting an attacker replay the whole attack path describe in figure \ref{scenario_fig}. This proof of concept is available online at \url{https://anonymous.4open.science/r/ursid-anon-23C1}. This includes the configuration file which resulted from the procedural refinement process, the code necessary to deploy the VMs alongside installation instructions, and a set of commands showing how to execute and reproduce the attack described on the resulting architecture. 

Figure \ref{fig:config_file} showcases a sample of this configuration file, related to machine Raccoon. This excerpt shows the operating system constraints that are required (here Debian at version 9.0), a software constraint (installing package sudo at version 1.8.2 in order to exploit CVE-2021-3156), a user constraint (creating user Bob and setting its password to a weak value) and finally two file constraints. One of those file constraints overwrites the bash history of the root user to leak passwords, and the other edits the sudoers file in order to exploit CVE-2021-3156 properly. "Source" here refers to the file path in the deployment machine, and "Destination" refers to the file path in the resulting machine. The content of the source files are automatically generated during the refinement process based on the chosen procedure. The full configuration file is also available online.

The deployment process for the whole architecture is as follows. First, all machines are generated and populated with the relevant users. Then for each machine, the operating system and software to install on it are collected and are then converted into instructions. This requires a library of operating system disk images and software installation instructions.
Unfortunately, the sheer diversity of architectures and different ways to install software requires a lot of those instructions to be written on a case by case basis. For instance, installing the latest sudo package on a Ubuntu architecture can be done with a single apt-get command, but installing a legacy sudo package (for instance one that is vulnerable to \cite{cve_report_sudo}) requires installing the package manually. Some inline commands may be ran during the installation process. 
The machines are then populated with files. This is either done by generating the files from scratch and copy pasting them into the relevant VM, using reference configuration files (that may be edited depending on the procedure), editing files that will naturally be available on the virtual machine (like system files or windows registry), or even copy pasting entire sample directories (useful for file corpuses). Finally, even after their deployment, some commands may need to be automatically ran on the virtual machines. For instance, this could detonate payloads or simulate user activity in our infrastructure, which may be relevant for some procedures or help with the credibility of our deception platform.

\begin{figure*}
\begin{minipage}[c]{0.8\linewidth}
\begin{Verbatim}
[{
	"Name": "Raccoon",
	"OS": {"Type": "debian:debian_linux", "Version": "9.0"},
	"Software": [{"Type": "sudo_project:sudo", "Version": "1.8.2"}],
	"Users": [{"Name": "Bob","Group": "Bob","Privilege": "User", 
		"Credentials": {"Type": "UnixUserAccount", "Value": "rainbow"}}],
	"Files": [{
		"Source": "./generated_files/Raccoon/bashrc_password_leak", 
		"Destination": "~/bash_history",
		"Permissions": "0600", "Modification": "WRITE",
		"Owner": "root", "Group": "root"
	},
	{
		"Source": "./generated_files/Raccoon/sudo_1.8.2_exploit",
		"Destination": "/etc/sudoers",
		"Permissions": "4111", "Modification": "APPEND",
		"Owner": "root", "Group": "root"}]
}]	
\end{Verbatim}
\end{minipage}
\caption{Excerpt of example configuration file.}\label{fig:config_file}
\end{figure*}

	\section{Conclusion and future works}
	In this work we presented a new formal representation of attack scenarios based on attack graphs. This representation uses the MITRE ATT\&CK nomenclature to label the attacker actions.
We demonstrated how to model an attack scenario on a technical level, then showcased a refinement process. This process lets us decline a unique scenario into a multitude of architecture configurations defined on a procedural level by computing procedure constraints. These architecture configurations may finally be deployed as functional honeynets vulnerable to the formalized attack scenario.

We have several plans to improve and expand on this project. First of all, we plan to cover more MITRE ATT\&CK techniques and enhance our procedure database for each of them. This would diversify the kinds of scenarios that may be deployed by our system, and make it easier to directly convert a technical report describing an attack scenario into an architecture using our process. 

We then plan to diversify the criterias that our backtracking process uses to pick its procedures. For instance the user could specify that procedures requiring a specific tool to be exploited have to be included in the scenario, in order to test whether the attacker is skilled at this particular tool. This would also include a way for the user to specify constraints for each device before the backtracking algorithm even takes place, by for instance specifying that the network has to be only comprised of Windows operating systems, in order to more accurately match the structure of an existing network.

We then aim to automatize the process of deploying additional decoy machines, by making it possible to add prefab machines instead of having to write every single attack position manually. Scenarios would thus be a combination of a written-by-hand attack path and automatically generated machine clusters to populate the network. This would help with the credibility of our deception platform by adding additional attack positions that are not mandatory for the attacker to win.

Finally we plan to deploy in the wild a scenario that has been generated from start to finish by our formalization and refinement process. We first aim to test such a deception platform by locally hired red teammers. We will then deploy it on a local university network (with security taken into account) rather than on online cloud platforms to make it look more appealing to potential attackers. We will analyze its performances in term of attractivity, average attacker skill and attacker retention, and compare it to other honeypots in the field.

	\bibliographystyle{ACM-Reference-Format}
	\bibliography{biblio}

	\appendix
	
	\section{Detailed constraints for six procedures.}

	\begin{figure*}
		\begin{center}
				\scalebox{0.8}{
					\begin{tabular}{ |c|c|c|c|c|c|c| } 
						\hline
						\textbf{Procedure} & \textbf{Machine} & \textbf{Secret Preconditions} &\textbf{ OS constraints} & \textbf{Account constraints} & 	\makecell[c]{\textbf{Software} \\ \textbf{constraints}} & \textbf{File constraints} \\
						\hline 
						\makecell[c]{CVE \\to  bypass \\ UAC} & Entry & \makecell[c]{\textit{Requires}: [] \\ \textit{Rewards}: []} & \makecell[c]{\textit{Type}: Windows \\ \textit{Version}: 10} & \makecell[c]{\textit{Username}: * \\ \textit{Group}: * \\ \textit{Privilege}: SuperUser \\ \textit{Credentials}: *} & \makecell[c]{\textit{Software}:\\ ixpdata:easyinstall \\ \textit{Version}: 6.2.13723 \\ \textit{Port}: *}  &  * \\
						\hline 
						\makecell[c]{Bad Sudo \\ Config} & Exit & \makecell[c]{\textit{Requires}: [] \\ \textit{Rewards}: []} & \makecell[c]{\textit{Type}: Linux \\ \textit{Version}: *} & * & * & \makecell[c]{\textit{Path}: ~/.sudoers \\ \textit{Permissions}: * \\ \textit{Content}: "Flawed Sudo \\ Config"} \\
						\hline 
						\makecell[c]{Password \\ in bash \\history} & Exit & \makecell[c]{\textit{Requires}: []\\ \textit{Rewards}: [Password, *]} & \makecell[c]{\textit{Type}: Linux \\ \textit{Version}: *} & \makecell[c]{\textit{Username}: * \\ \textit{Group}: * \\ \textit{Privilege}: SuperUser \\ \textit{Credentials}: *} & * &  \makecell[c]{\textit{Path}: ~/bash\_history \\ \textit{Permissions}: * \\ \textit{Content}: "Bashrc Password Leak"}\\
						\hline 
						\makecell[c]{Password in \\ text file} & Exit & \makecell[c]{\textit{Requires}: []\\ \textit{Rewards}: [Password, *]} & * & * & * &  \makecell[c]{\textit{Path}: ~/reminder.txt \\ \textit{Permissions}: * \\ \textit{Content}: "txt password leak}\\
						\hline 
						\makecell[c]{SSH Access\\from key} & Exit & \makecell[c]{\textit{Requires}: [SSH key, 1] \\ \textit{Rewards}: []} & \makecell[c]{\textit{Type}: Linux \\ \textit{Version}: *} & - &  \makecell[c]{\textit{Software}: ssh:ssh \\ \textit{Version}: * \\ \textit{Port}: 22}  &  \makecell[c]{\textit{Path}: ~/.ssh/authorized\_keys \\ \textit{Permissions}: 644 \\ \textit{Content}: "Add public SSH key}\\
						\hline 
						\makecell[c]{RDP \\ Access from \\ password} & Exit & \makecell[c]{\textit{Requires}: [Password, 1] \\ \textit{Rewards}: []} & \makecell[c]{\textit{Type}: Windows \\ \textit{Version}: *} & \makecell[c]{\textit{Username}: * \\ \textit{Group}: * \\ \textit{Privilege}: * \\ \textit{Credentials}: \\Password from Secret}  &  \makecell[c]{\textit{Software}: \\ microsoft:\\ remote\_desktop \\ \textit{Version}: * \\ \textit{Port}: 3389}  &  \makecell[c]{\textit{Path}:\\ "HKEY\_LOCAL\_MACHINE\textbackslash\textbackslash \\ SYSTEM \textbackslash \textbackslash CurrentControlSet \\ \textbackslash \textbackslash Control\textbackslash \textbackslash Terminal Server" \\ \textit{Permissions}: * \\ \textit{Content}: "Windows RDP Registry}\\
						\hline 
					\end{tabular}
				}
			
			\caption{Detailed constraints for 6 procedures}\label{fig:new_table_example}
		\end{center}
	\end{figure*}

\end{document}